\documentclass[a4paper,11pt]{article}
\usepackage{jinstpub} 
\usepackage{lineno}

\title{\boldmath Setups for eliminating static charge of the ATLAS18 strip sensors}

\author[a,1]{P. Federi\v{c}ov\'{a},}
\author[b]{A.Affolder,}
\author[d]{G.A. Beck,}
\author[d]{A. J. Bevan,}
\author[d]{Z. Chen,}
\author[d]{I. Dawson,}
\author[b]{A. Deshmukh,}
\author[b]{A. Dowling,}
\author[b]{V. Fadeyev,}
\author[c,e]{J. Fernandez-Tejero,}
\author[c,e]{A. Fournier,}
\author[b]{N. Gonzalez,}
\author[f]{L. Hommels,}
\author[g]{C. Jessiman,}
\author[b]{S. Kachiguin,}
\author[g]{Ch. Klein,}
\author[g]{T. Koffas,}
\author[a]{J. Kroll,}
\author[a]{V. Lato\v{n}ov\'{a},}
\author[a]{M. Mike\v{s}t\'{i}kov\'{a},}
\author[d]{P. S. Miyagawa,}
\author[c,e]{S. O‘Toole,}
\author[b]{Q. Paddock,}
\author[c,e]{L. Poley,}
\author[g]{E. Staats,}
\author[c,e]{B. Stelzer,}
\author[a]{P. Tůma,}
\author[h]{M. Ullan,}
\author[i]{Y. Unno,}
\author[d]{S. C. Zenz.}

\affiliation[a]{Institute of Physics, Czech Academy of Sciences,\\
Na Slovance 2, 18200 Prague 8, the Czech Republic}
\affiliation[b]{Santa Cruz Institute for Particle Physics (SCIPP),\\
University of California, Santa Cruz, CA 95064, USA}
\affiliation[c]{Department of Physics, Simon Fraser University,\\
8888 University Drive, Burnaby, B.C. V5A 1S6, Canada}
\affiliation[d]{Department of Physics and Astronomy,\\
Queen Mary University of London, London, E1 4NS, UK}
\affiliation[e]{TRIUMF/SFU,\\
4004 Wesbrook Mall, Vancouver, B.C. V6T 2A3, Canada}
\affiliation[f]{Cavendish Laboratory, University of Cambridge,\\
JJ Thomson Avenue, Cambridge CB3 0HE, United Kingdom}
\affiliation[g]{Physics Department, Carleton University,\\
1125 Colonel By Drive, Ottawa, Ontario, K1S 5B6,  Canada}
\affiliation[h]{Instituto de Microelectrónica de Barcelona (IMB-CNM, CSIC),\\
Campus UAB-Bellaterra, 08193  Barcelona, Spain}
\affiliation[i]{Institute of Particle and Nuclear Study, High Energy Accelerator Research Organization (KEK),\\
1-1 Oho, Tsukuba, Ibaraki 305-0801, Japan}

\emailAdd{pavla.federicova@cern.ch}

\abstract{Construction of the new all-silicon Inner Tracker (ITk), developed by the ATLAS collaboration to be
able to track charged particles produced at the High-Luminosity LHC, started in 2020 and is expected to
continue till 2028. The ITk detector will include 18,000 highly segmented and radiation hard n+-in-p
silicon strip sensors (ATLAS18), which are being manufactured by Hamamatsu Photonics. Mechanical and electrical characteristics of produced sensors are measured upon their delivery at several institutes participating in a complex Quality Control (QC) program. The QC tests performed on each individual sensor check the overall integrity and quality of the sensor. 
During the QC testing of ATLAS18 strip sensors, an increased number of sensors that
failed the electrical tests was observed. In particular, IV measurements indicated an early breakdown,
while large areas containing several tens or hundreds of neighbouring strips with low interstrip isolation
were identified by the Full strip tests, and leakage current instabilities were measured in a long-term leakage current stability setup. Moreover, a high surface electrostatic charge reaching a level of several hundreds of volts per inch was measured on a large number of sensors and on the plastic sheets, which
mechanically protect these sensors in their paper envelopes. Accumulated data indicates a clear
correlation between observed electrical failures and the sensor charge-up.
To mitigate the above-described issues, the QC testing sites significantly modified the sensor handling
procedures and introduced sensor recovery techniques based on irradiation of the sensor surface with UV
light or application of intensive flows of ionized gas. In this presentation, we will describe the setups
implemented by the QC testing sites to treat silicon strip sensors affected by static charge and
evaluate the effectiveness of these setups in terms of improvement of the sensor performance. \footnote{Copyright 2023 CERN for the benefit of the ATLAS Collaboration. Reproduction of this article or parts of it is allowed as specified in the CC-BY-4.0 license.}
}


\begin{document}
\maketitle
\flushbottom

\section{Introduction}
\label{sec:intro}

The upgrade of the Large Hadron Collider (LHC) into the High-Luminosity LHC (HL-LHC) requires the replacement of the current ATLAS Inner Detector with a new all-silicon tracker (ITk) containing new types of silicon pixel and strip sensors~\cite{1}. The ITk strip detector will expand the silicon area from 60 m$^{2}$ to 165 m$^{2}$, while substantially improving radiation hardness and signal processing speed. The production of a total of 21 000 ITk strip sensors, including spares to the $~$18 000 sensors to be used in the experiment, comprising 2 types of barrel and 6 types of endcap sensors\cite{2}, has started in Hamamatsu Photonics K.K in 2021 and is scheduled for completion in 2025.

To ensure the overall integrity and quality of each individual sensor, QC tests~\cite{3} are performed at various participating institutes. These QC tests include Visual inspection and capture, Metrology to assess sensor bow and thickness, IV and CV measurements, evaluation of the Long-term stability of leakage current, and comprehensive Full strip testing.

\section{Issues and Solutions}

During the QC testing of ATLAS18 strip sensors, some QC sites (Queen Mary University of London (QMUL), Carleton University, University of Cambridge, Institute of Physics in Prague (FZU), Santa Cruz Institute for Particle Physics (SCIPP), Simon Fraser University (SFU), and Canada's particle accelerator center TRIUMF) recorded an increase in the number of sensors that failed electrical tests. Specific issues observed include:

\begin{enumerate}
    \item  \textbf{IV measurements indicating early breakdown}: Sensors experience breakdown at low applied bias voltages, well below the required 500 V.
    \item  \textbf{Low interstrip isolation} in neighboring strips: Low interstrip isolation between neighboring strips: Areas of tens or hundreds of neighboring strips with a low resistance of polysilicon bias resistors were revealed by the Full strip tests, indicating low interstrip isolation in these regions~\cite{4}.
    \item \textbf{Significant variation in leakage current}: Measurements of the long-term stability of the sensor leakage current showed significant variation of this current over a 48-hour long testing period.
\end{enumerate}
    
\noindent Accumulated data suggested a correlation between the observed electrical failures of tested sensors and a high static charge measured on the surface of these sensors and their protecting plastic sheets. The presence of a high surface charge increases the risk of "Local trapped charge" events, which may have contributed to the failures of electrical tests of studied silicon strip sensors.
\\
\noindent These issues were reported to the vendor for a possible solution. To mitigate them on the ATLAS side, the QC sites implemented several measures:
\begin{enumerate}
    \item \textbf{Sensor handling procedure modification}: The sensor handling procedures were modified to completely eliminate or at least significantly reduce the risk of "Local trapped charge" events during QC testing of production sensors - proper grounding of laboratory equipment and measurement devices was implemented to prevent the build-up of static charge, while the grounding of laboratory workers should avoid the appearance of "Local trapped charge" events on tested samples.
    \item \textbf{Ionizing gas treatment}: Intensive flows of ionized gas (air, nitrogen) are applied to neutralize any static charge accumulated on the surface of sensors or their protective shields. The ionized gas is also used as a recovery method as it helps to dissipate the charge stuck in the sensor surface layers.
    \item \textbf{UV light irradiation}: UV light irradiation is used as a sensor recovery method - UV light exposure dissipates a charge, which was stuck in the surface layers of the sensor due to the "Local trapped charge" event.
    \item \textbf{High-temperature exposure}: Subjecting the sensors to high-temperature exposure is used to reset or stabilize their electrical properties - controlled high-temperature treatment removes the defects caused by the "Local trapped charge" events.
\end{enumerate}


\section{Electrostatic charge and "Local trapped charge" events}

\begin{figure}[htbp]
\centering
\includegraphics[width=0.4\textwidth]{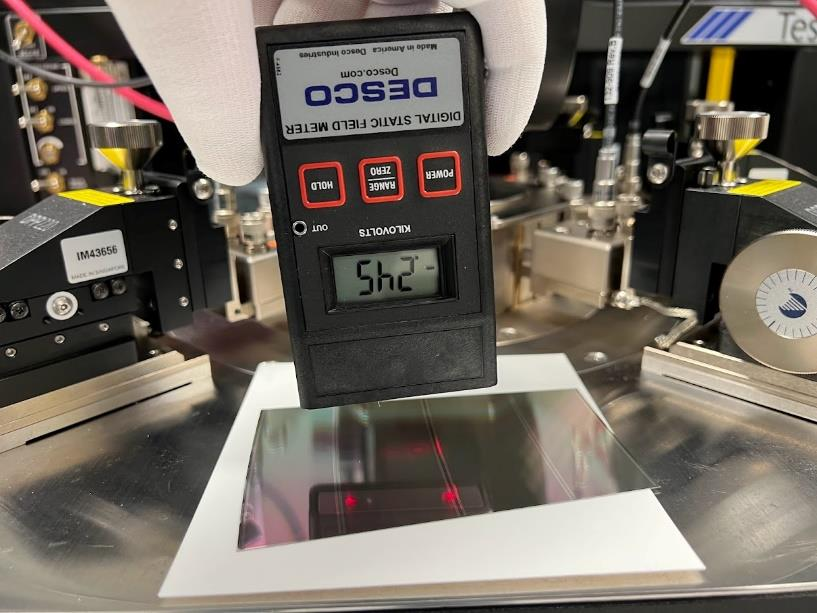}
\qquad
\includegraphics[width=0.4\textwidth]{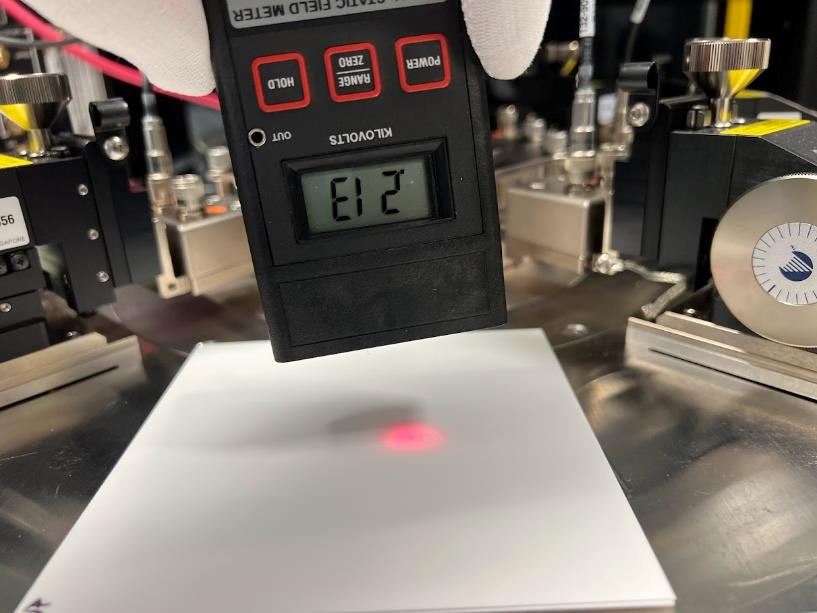}
\caption{Measurement of static charge on sensors and sheets with an electrostatic fieldmeter in FZU.\label{1}}
\end{figure}

\begin{figure}[htbp]
\centering
\includegraphics[width=.77\textwidth]{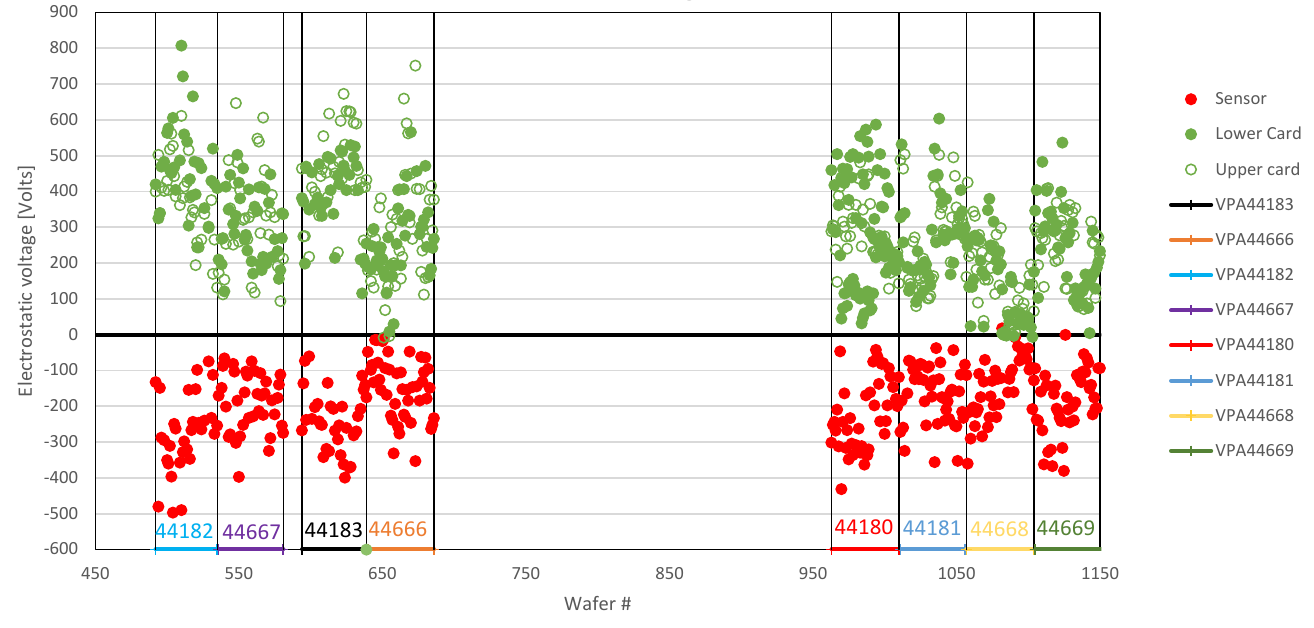}
\qquad
\includegraphics[width=.8\textwidth]{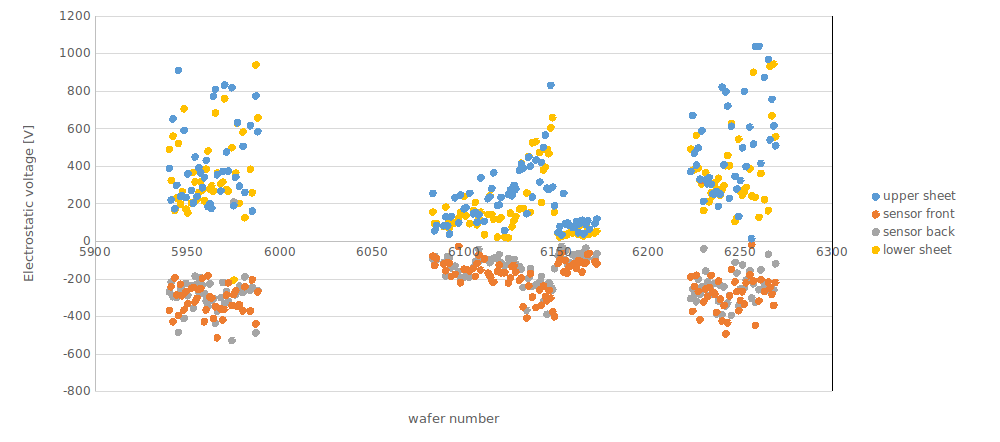}
\caption{Electrostatic charge measured on the surface of sensors and their sheets in FZU (top) and in QMUL (bottom).\label{2}}
\end{figure}

The QC sites have measured a significant electrostatic charge present on the surface of a considerable number of sensors and their plastic sheets, see Figure~\ref{1} for measurement in FZU. The plastic sheets are used to provide mechanical protection for the sensors within paper envelopes. The sensors themselves have a negative static charge, while the plastic sheets carry a positive charge. In both cases, the electrostatic charge can reach the values of several hundreds of volts per inch, see Figure~\ref{2}.

The electrostatic charge, generated by various means, including the rubbing of plastic sheets against sensors within paper envelopes, facilitates the occurrence of "Local trapped charge" events. These events subsequently lead to failures in electrical QC tests at several QC sites.

\subsection*{Process of "Local trapped charge" events}

A "Local trapped charge" event can occur in various scenarios during the manufacturing of sensors, their shipment, and further handling. These events can result in the accumulation of charge in the surface layers of the sensor (if the "Local trapped charge" event was caused by touching the sensor surface by the vacuum stencil or other tool, the charge is stuck at the point of contact). The electrostatic charge accumulated on the sensor surface can lead to the loss of interstrip isolation provided by the p-stop implant, while its presence close to the edge and bias ring of the sensor causes instability of the leakage current or the sensor breakdown at low bias voltages. {The p-stop is not sufficient to compensate the electrostatic charge, because its role and the surface charge influence are competing effects. For example, the strip isolation goes down with an ionizing dose that raises the surface charge~\cite{5}. Originally the p-stop concentration was chosen on the basis of good isolation and other properties such as strip capacitance and breakdown performance~\cite{6}.}

\section{Recovery setups}

 In case the "Local trapped charge" event occurs on the sensor, this sensor can be permanently damaged. However, it is shown that the majority of affected sensors can be recovered by the application of specific recovery procedures, such as exposure to the intensive flow of an ionized gas, UV light irradiation, or a high-temperature treatment. All mentioned methods effectively dissipate the charges stuck in the surface 
 of the SiO$_{2}$ passivation layers or in the interface between the passivation layer and the metal or silicon, by inducing additional charges, and thus stabilize the electrical characteristics of the affected sensors. 
 
 \vspace{0.5cm}
\noindent Recovery methods used by the QC sites:
\begin{enumerate}
    \item \textbf{UV-A }(315-400 nm wavelength) light setups with typical exposure between 2 and 8 hours,
    \item \textbf{UV-C} (100-280 nm wavelength) light setups with typical exposure of 60 seconds,
    \item \textbf{ionizing air blowers} with a typical exposure of up to 30 minutes,
    \item \textbf{high-temperature exposure} ("baking") of sensors in a bake-out oven for more than\\ 16 hours at 150 ºC. 
\end{enumerate}

\begin{figure}[htbp]
\centering
\includegraphics[width=.245\textwidth]{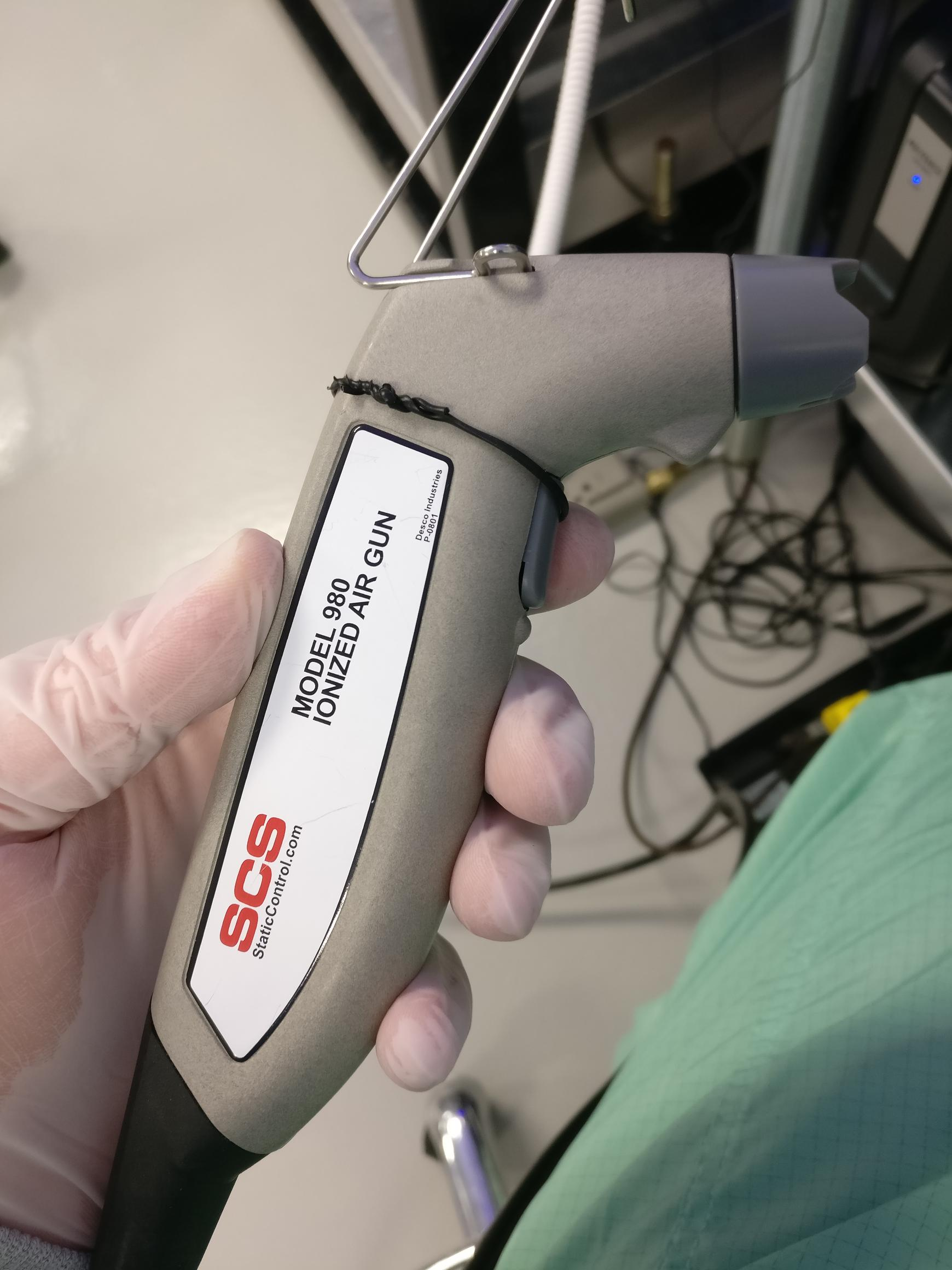}
\includegraphics[width=.3\textwidth]{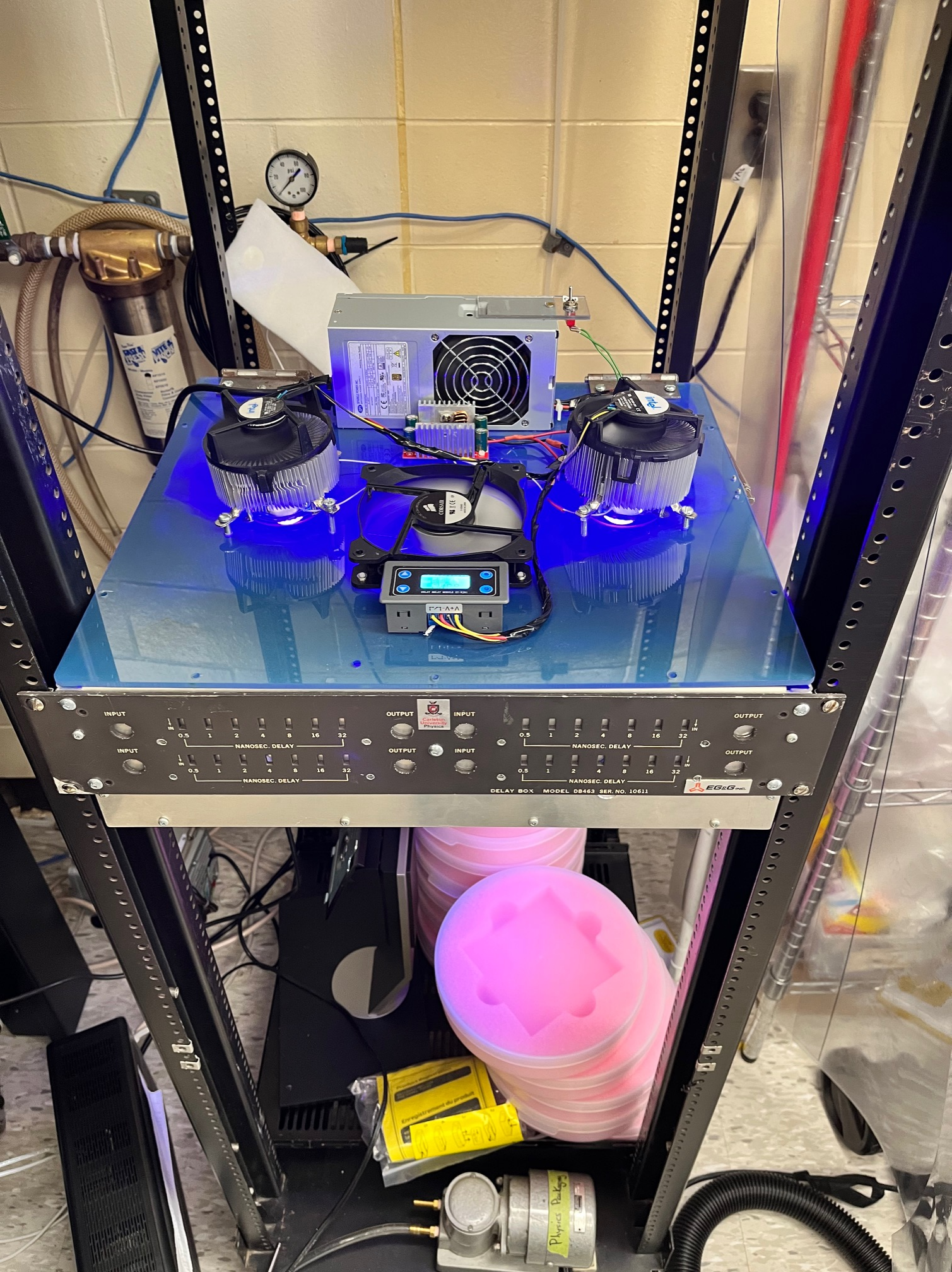}
\includegraphics[width=.3\textwidth]{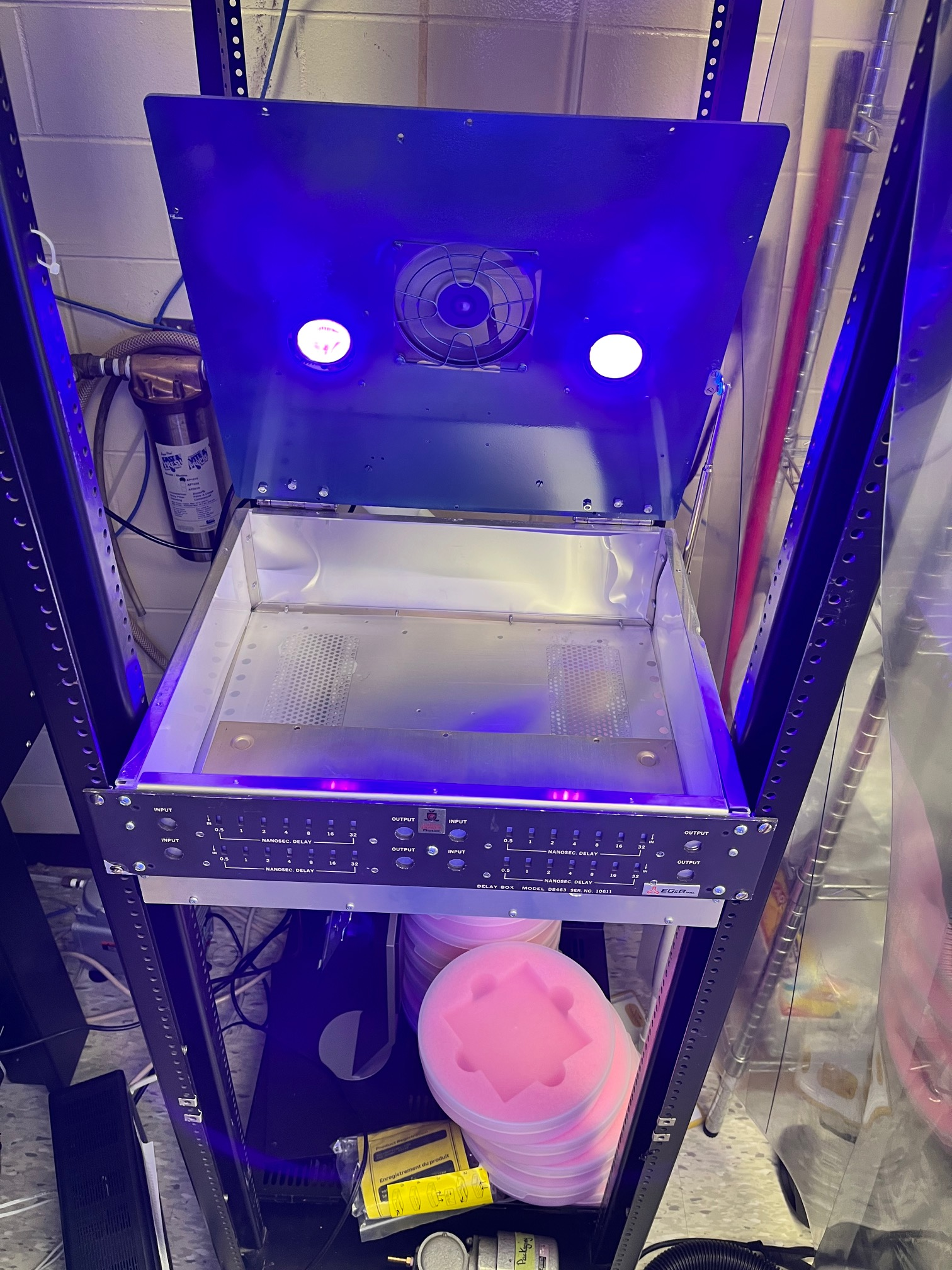}
\caption{Carleton ion gun (left) and UV-A setup (middle and right).\label{3}}
\end{figure}

\begin{figure}[htbp]
\centering
\includegraphics[width=.41\textwidth]{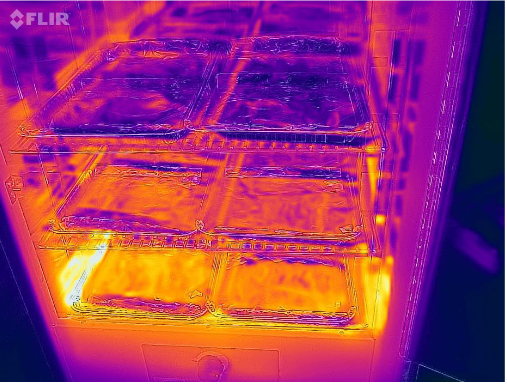}
\includegraphics[width=.4\textwidth]{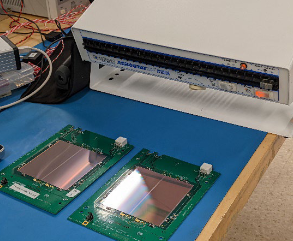}
\caption{SCIPP baking oven (left) and ion blower setup (right).\label{4}}
\end{figure}

\vspace{0.5cm}
\noindent Sensor recovery setups used at the individual ATLAS ITk strip sensor QC sites include (ordered as they are used in the individual QC sites):
\begin{itemize}
    \item Carleton: UV-C, UV-A, ion-gun (Figure~\ref{3}),
    \item FZU: UV-A, ion-gun, ion-blower,
    \item SCIPP: UV-A, baking, ion-blower (Figure~\ref{4}),
    \item QMUL: ion-blower, UV-C,
    \item Cambridge: ion-blower,
    \item TRIUMF/SFU: UV-A.
\end{itemize}

\begin{figure}[htbp]
\centering
\includegraphics[width=1\textwidth]{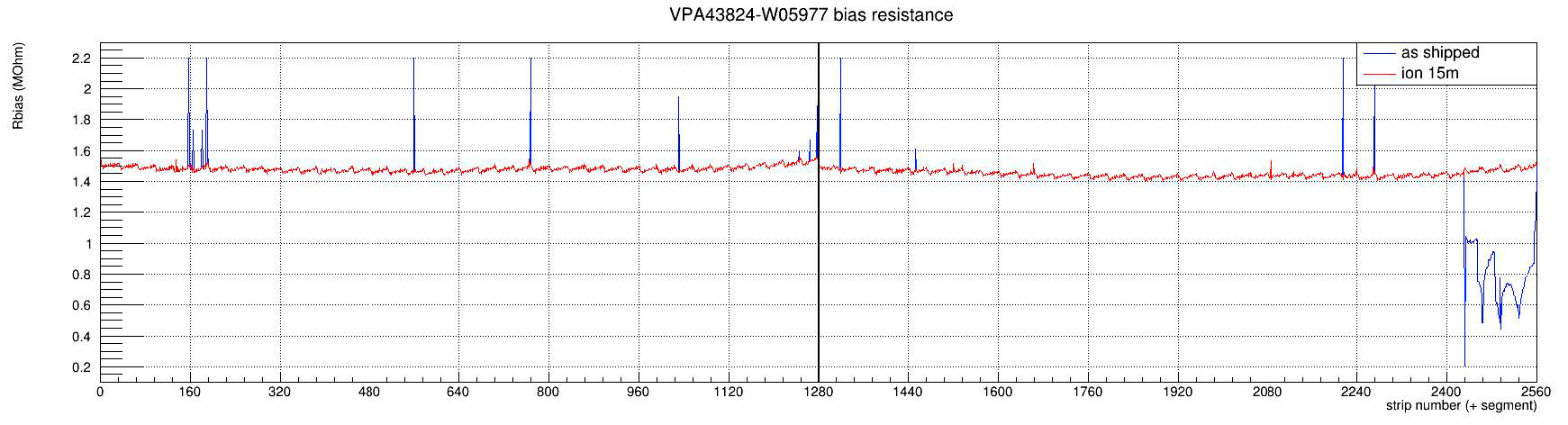}
\caption{The Full strip test performed on the sensor before and after the ion-blower treatment in QMUL. The resistance of the polysilicon bias resistor was measured on all strips of the sensor VPA43824-W05977. The area of~160 neighboring strips (around strip 2500) with low interstrip isolation was recovered by the ion blower treatment.\label{5}}
\end{figure}

\begin{figure}[htbp]
\centering
\includegraphics[width=0.47\textwidth]{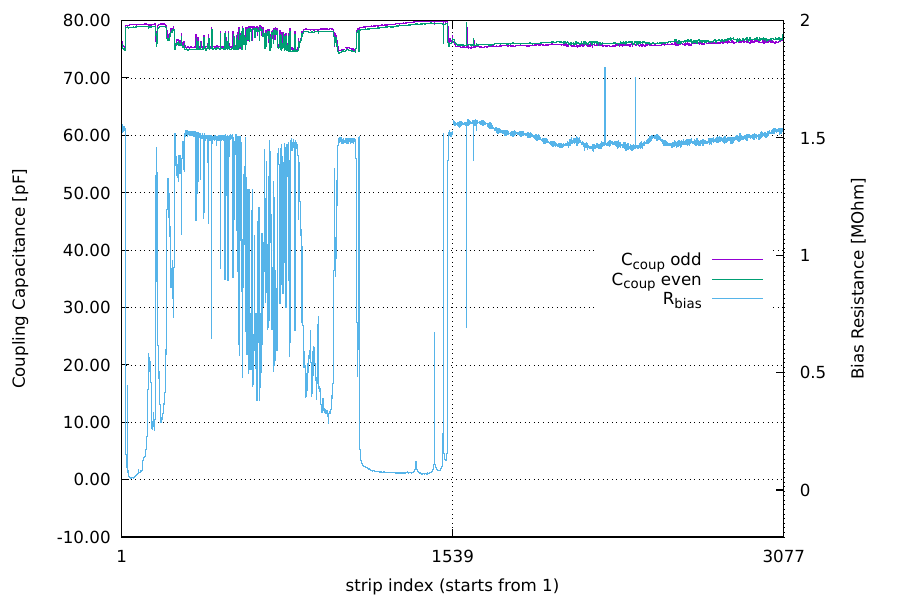}
\includegraphics[width=0.47\textwidth]{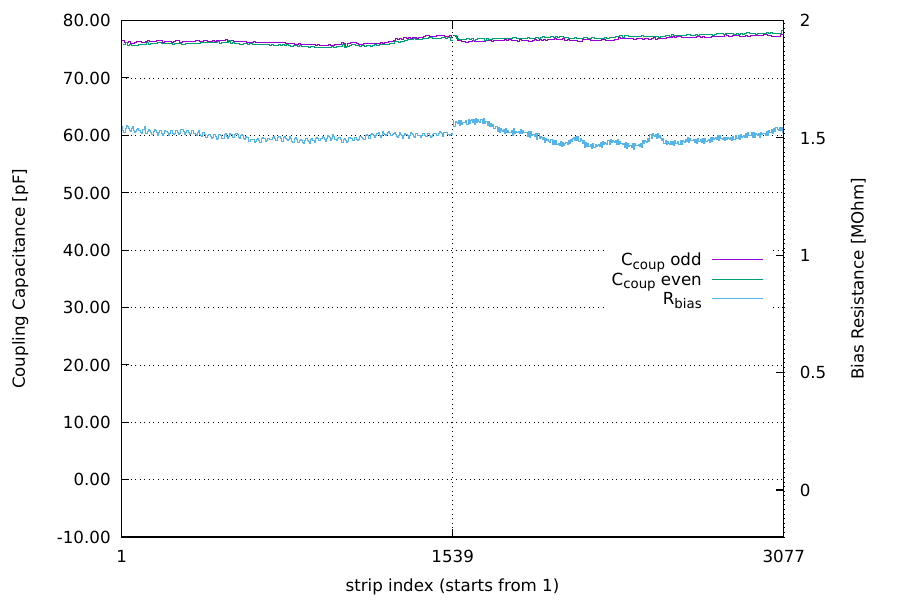}
\caption{The Full strip test performed before (left) and after the UV-A treatment (right) in FZU. The resistance of the polysilicon bias resistor and coupling capacitance was measured on all strips of the sensor VPA38205-W00189. The area of$~\sim$1500 strips with low interstrip isolation was recovered.\label{6}}
\end{figure}

\begin{figure}[htbp]
\centering
\includegraphics[width=0.49\textwidth]{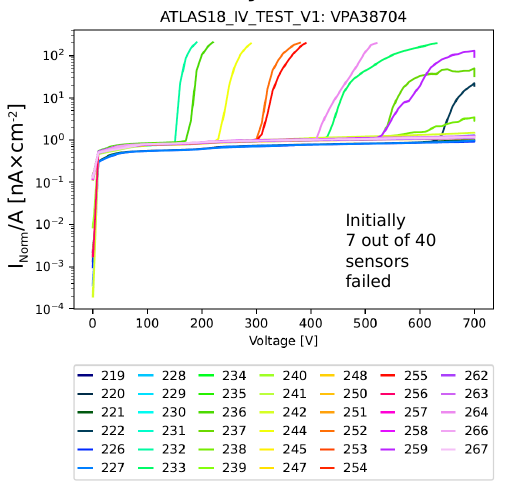}
\includegraphics[width=.49\textwidth]{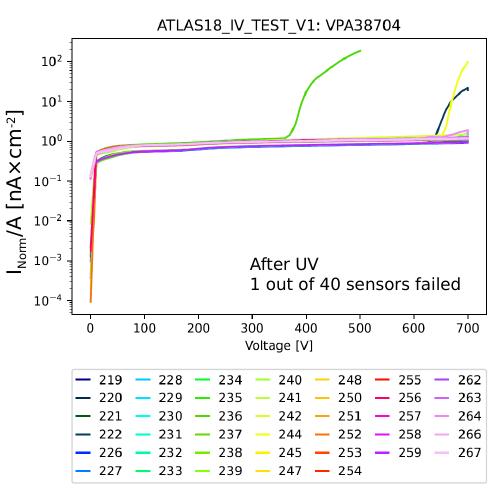}
\caption{The IV measurements for the batch VPA38704 before and after the UV-A treatment in FZU. The current normalized to the surface of a sensor in the dependence of high voltage was measured for all sensors in the aforementioned batch. The majority of sensors with lower breakdown voltage were fully recovered.\label{7}}
\end{figure}

\noindent Application of the implemented recovery methods on the sensors affected by the "Local trapped charge" events is typically very effective. In some cases, several iterations are needed to fully recover the electrical characteristics of these sensors. The recovery of the low interstrip isolation by the intensive flow of ionized air is shown in Figure~\ref{5}, while the application of the UV-A irradiation is illustrated in Figure~\ref{6}. A significant increase in the breakdown voltage by the UV-A irradiation is represented by Figure~\ref{7}.

{However, the static charge can be generated at any point during the assembly process. Ionizing blowers have been effectively used on modules and staves to remove the static influence. It should also be noted that there is no direct contact with the sensors' surface after staves and petals are built. }

\section{Summary}

A significant increase in failed electrical QC tests was observed by the institutes participating in the QC testing of production ATLAS ITk silicon strip sensors. The accumulated data indicated a clear correlation between the electrical failures of studied sensors and the level of static charge measured on their surface. To address individual issues related to the static charge, the QC testing sites modified their handling procedures. Additionally, they successfully employed various sensor recovery methods, including exposure to UV-A/UV-C light or an intensive flow of ionized gas, as well as sensor baking. These measures have proven to be very effective in mitigating the impact of "Local trapped charge" events and improving the overall performance of the sensors.

\acknowledgments

This work was supported by the Ministry of Education, Youth and Sports of the Czech Republic coming from the projects LTT17018 Inter-Excellence and LM2018104 CERN-CZ, 
by Charles University grant GAUK 942119, by the US Department of Energy, grant DE-SC0010107, by the Canada Foundation for Innovation and the Natural Science and Engineering Research Council of Canada, and by the Spanish R\&D grant PID2021-126327OB-C22, funded by MCIN/ AEI/10.13039/501100011033 / FEDER, UE.




\end{document}